 \documentclass[11pt,journal,a4paper,draftcls,oneside,onecolumn]{IEEEtran}

\usepackage{url}
\usepackage{subfigure}
\usepackage{amsmath,amssymb}	
\usepackage{graphicx,cite,url,array}
\usepackage[outdir=./]{epstopdf}
\usepackage[usenames]{color}

\usepackage{trackchanges}
\usepackage{soul}
\addeditor{Yu}
\usepackage{wrapfig}
\usepackage{amssymb}
\usepackage{ulem}
\usepackage{url}

\newcommand{\figref}[1]{Fig.~\ref{#1}}

\begin{document}
\title{An Immunology-Inspired Network Security Architecture}
\author{Quan~Yu,~\IEEEmembership{Senior Member,~IEEE},~Jing~Ren,~Jiyan~Zhang,~Siyang~Liu,\\ ~Yinjin~Fu,~Ying~Li,~Linru~Ma,~Jian~Jing,~Wei~Zhang,~\IEEEmembership{Fellow,~IEEE}
\thanks{
Q. Yu is with Peng Cheng Laboratory, Shenzhen, China (email: yuq@pcl.ac.cn). 
J. Ren is with School of Information and Communication Engineering, University of Electronic Science and Technology of China, Chengdu, China, (email: 
renjing@uestc.edu.cn).
J. Zhang is with Beijing Institute of Brain Sciences, Beijing, China (email: zhangjy@nic.bmi.ac.cn)
S. Liu is with Department of Electronic and Engineering, Shanghai Jiao Tong University, Shanghai, China (email: saberkingdom@sjtu.edu.cn)
Y. Fu is with Peng Cheng Laboratory, Shenzhen, China (email: fuyj@pcl.ac.cn). 
Y. Li is with Peng Cheng Laboratory, Shenzhen, China (email:  liy02@pcl.ac.cn). 
L. Ma is with System Engineering Research Institute, Beijing, China (email: malinru@163.com)
J. Jing is with State Key Laboratory of Pharmaceutical Biotechnology, School of Life Sciences, Nanjing University, Nanjing, China (email: jingj01@live.com))
}
\thanks{
Corresponding author:  W. Zhang is with Peng Cheng Laboratory, Shenzhen, China, and School of Electrical Engineering \& Telecommunications, University of New South Wales, Sydney, Australia (email: w.zhang@unsw.edu.au).}
%\thanks{Corresponding author: W. Zhang}
}

\maketitle

\begin{abstract}
The coming 5G networks have been enabling the creation of a wide variety of new services and applications which demand a new network security architecture.  Immunology is the study of the immune system in vertebrates (including humans) which protects us from infection through various lines of defence. By studying the resemblance between the immune system and network security system, we acquire some inspirations from immunology and distill some guidelines for the design of network security architecture. We present a philosophical design principle, that is maintaining the balance between security and availability. Then, we derive two methodological principles: 1) achieving situation-awareness and fast response through community cooperation among heterogeneous nodes, and 2) Enhancing defense capability through consistently contesting with invaders in a real environment and actively mutating/evolving attack strategies.
We also present a reference architecture designed based on the principles.

\end{abstract}

\newpage
\section*{Introduction}
\label{sec.Introduction}

Internet of everything (IoE) has been considered as the future of Internet that connects  people, processes, data and things. 
% Over recent decades, the Internet has fundamentally changed the way we access and exchange information
%At the end of 2017, there are 18 billion devices, most of which are IoT devices, connected to the Internet, which means 2.4 networked devices per person \cite{forecast2019cisco}.  
It was recently reported that the number of connected devices would reach 50 billion by 2020 and 500 billion by 2030 \cite{forecast2019cisco}. 
 However, {the high flexibility and rapid provisioning that Internet offers pose great challenges to network security systems.} 
The reason is that the Internet is initially designed based on the assumption of ``a group of mutually trusting users attached to a transparent network" \cite{blumenthal2001rethinking}, and security is one of the lowest priorities for the Internet's designers. 
It results in various severe security issues, which can be roughly classified into two categories.
\begin{itemize}
	\item \textbf{Information Confidentiality and Message Integrity:} 
	Due to the openness of the Internet, malicious users can steal legal users' information by wiretapping or masquerading as someone legal users trust. Then, they can deduce the message meanings (confidentiality) and even alter the content (integrity). This kind of security breach is difficult to detect as there is no invasion operation	and can only be solved by cryptographic techniques and authentication. 
	\item \textbf{Operational Security:} It refers to security issues when unauthorized users (from outside and inside) try to invade computing systems (including computers and other smart devices) to control these systems, steal legal users' information, launch attacks, extort money, or perform other malicious activities \cite{james2017computer}. 
	The defining feature of this kind of security issue is that the invaders perform some operations to invade the system. Thus, computer viruses, worms, and DDoS attacks belong to operational security issues. 
\end{itemize}

Here, we focus on the operational security issue which is  harmful and has caused significant financial losses \cite{US2019report}. 
%Report from the Council of Economic Advisers in the United States government estimated that the cost of the malicious cyber activity (such as DDoS, ransomware, virus) for US firms is \$498 million per adverse cyber event in 2018 \cite{US2019report}. 
Furthermore, IoT devices pose further challenges to network security as these small  things/devices have limited power, computing capacities and memories \cite{khan2018iot}. 
%In this paper, we aim to deal with this kind of security issue. 
 {To deal with the operational security issues, various methods have been proposed.}   
Some security strategies, like firewalls, virus scans, and intrusion detection, aim  to build  walls to block malicious attackers.  
Other strategies, like honeypots or honeynet \cite{spitzner2003honeynet}, cyber range \cite{ferguson2014national}, moving target network defense \cite{jajodia2011moving}, trusted computing \cite{mitchell2005trusted}, and mimic defense \cite{hu2017mimic},   provide active defense capabilities.  
These methods work well under some circumstances but also have limitations, such as 
\begin{itemize}
\item failing to counter unknown attacks, e.g., virus scans, and intrusion detection;
\item deploying defense solutions without collaboration, e.g., honeypots;
\item sacrificing universality and openness in exchange for security by employing dedicated-designed proprietary systems, e.g., trusted computing;
\item trading availability for security, e.g., mimic defense. 
\end{itemize}

\begin{figure*}[htb]
\centering
\includegraphics[width = 5in]{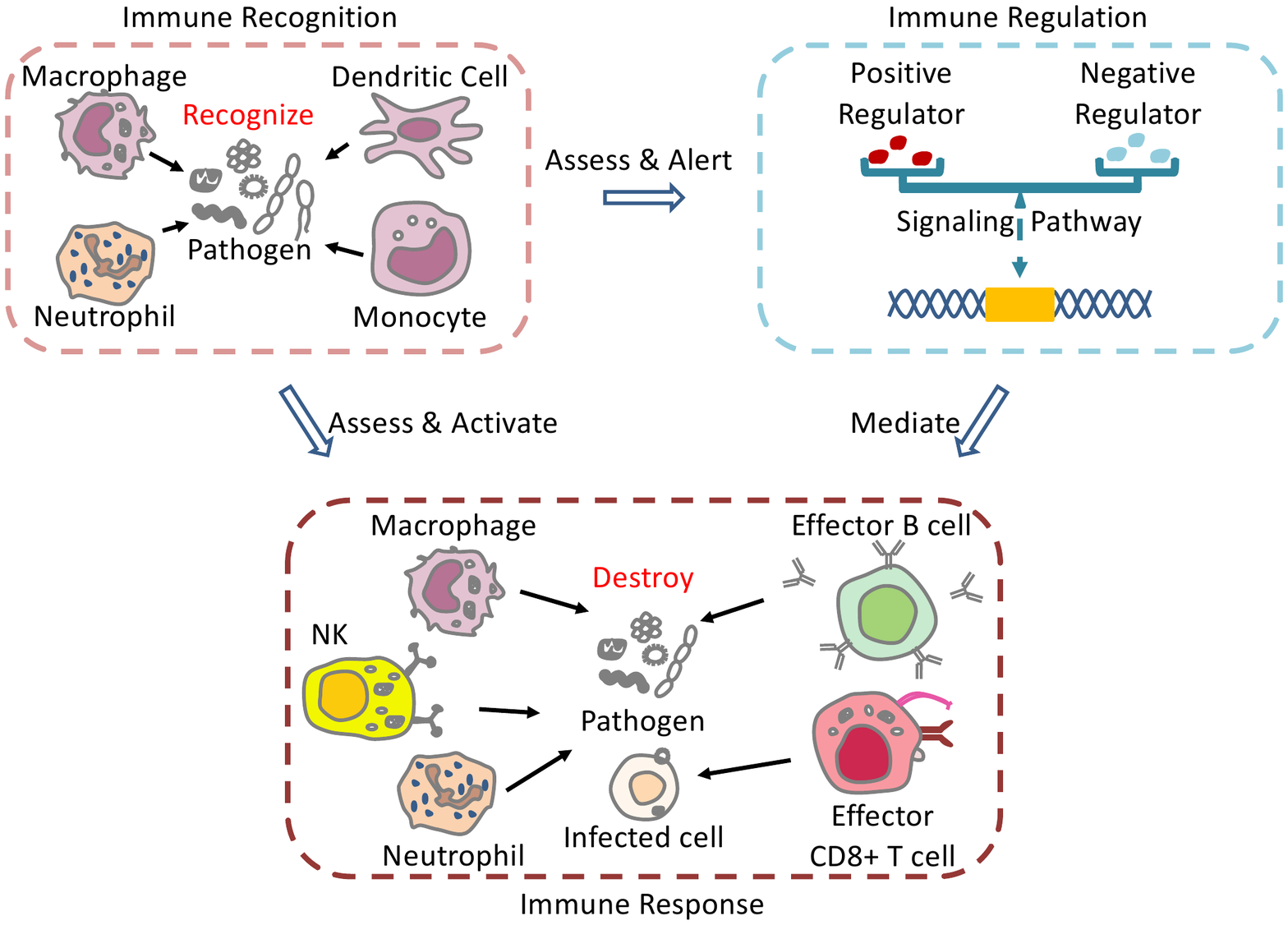}
\caption{The Primary Functions of the Biological Immune System in Vertebrates}
\label{fig:Functions}
\end{figure*}

To address these limitations, we need to rethink the systematic design of network security architecture.  
%Those limitations 
 %The root of these defects is that they try to figure out the security issues by patching the existing architecture, which cannot work when facing uncertainty. 
%What we need is a security architecture designed based on appropriate design principles for uncertain environments, which is an inflexible task and needs to take a lesson from other disciplines. 
 In this article, by studying the resemblance between the immune system in vertebrates and network security system, we  acquire some inspirations from immunology and distill some guidelines for the design of network security architecture.  We observe that 
the warfare between the biological immune system and combating pathogens, which grows in complexity and gradually vary their means of attack, is very similar to the battle between network security mechanisms and invaders.  Based on this observation, we present a philosophical design principle, that is maintaining the balance between security and availability and we give two methodological principles: 1) achieving situation-awareness and fast response through community cooperation among heterogeneous nodes, and 2) Enhancing defense capability through consistently contesting with invaders in a real environment and actively mutating/evolving attack strategies.
%We also present a reference architecture designed based on the principles. 
%We acquire five inspirations from the biological immune system. Then, we refine these inspirations into three design principles for security architecture:
%\begin{itemize}
%\item \textbf{One philosophy principle}:
%	\begin{itemize}
%		\item Keeping the balance between security and availability;
%	\end{itemize}
%\item \textbf{Two methodological principles}:
%	\begin{itemize}
%		\item Achieving situation-awareness and fast response through community cooperation among heterogeneous nodes;
%		\item Learning adversarial attacks in a controllable real environment.
%	\end{itemize}
%\end{itemize}

\iffalse
\begin{figure*}[!t]
\centering
\includegraphics[width = 7in]{fig/functions2.pdf}
\caption{The functions of the Immune system - alternative one}
\label{fig:Functions2}
\end{figure*}
\fi

The rest of this article is organized as follows. 
We first present an overview of the biological immunology and extract five inspirations. 
We then derive corresponding design principles of network security architecture from these inspirations and present a reference architecture.  
Finally, we draw conclusion and discuss future works.

 \section*{Inspirations from Biological Immunology for Network Security}
\label{sec.inspirations}

In this section, we firstly give a brief overview of the immune system in vertebrates. Then, we discuss the basic processes of the immune system, followed by some advanced features. Finally, we present the inspirations learned  from the immune system.

\subsection{The Overview of Biological Immune System}
The immune system has evolved billions of years to a complex host defense system which protects the human body from pathogens. 
 The primary functions provided by the biological immune system are illustrated in \figref{fig:Functions}.

The immune system continuously prevents pathogens from entering the body. Once it fails, the immune system tries to recognize these ``foreign'' invaders and assess the threat of the invasion, which is the process of \textbf{immune recognition}. 
If the invader is harmful, it activates the \textbf{immune response} to suppress the reproduction of pathogens, limit their harmful effects, and eventually destroy them or block them.

During this process, the immune system keeps monitoring the entire state of the body to maintain the balance between an effective immune response and an overaction, which is called \textbf{immune regulation} \cite{abbas2014cellular}. 
In practice, the immune system assesses the active level of invaders and the consequence of the immune response. 
If the invaders are active, through secreting positive regulators, the immune system can activate signaling pathways accelerating the immune response. 
Meanwhile, it also excretes negative regulators to restrict the immune response to prevent the body from entering a state of autoimmune disorders, hypersensitivities, or chronic inflammatory\cite{owen2013kuby}. 
This counterbalance is crucial for the homeostasis of an immune response\cite{immuneRegulation2007}.

\begin{figure*}[!t]
\centering
\includegraphics[width = 5in]{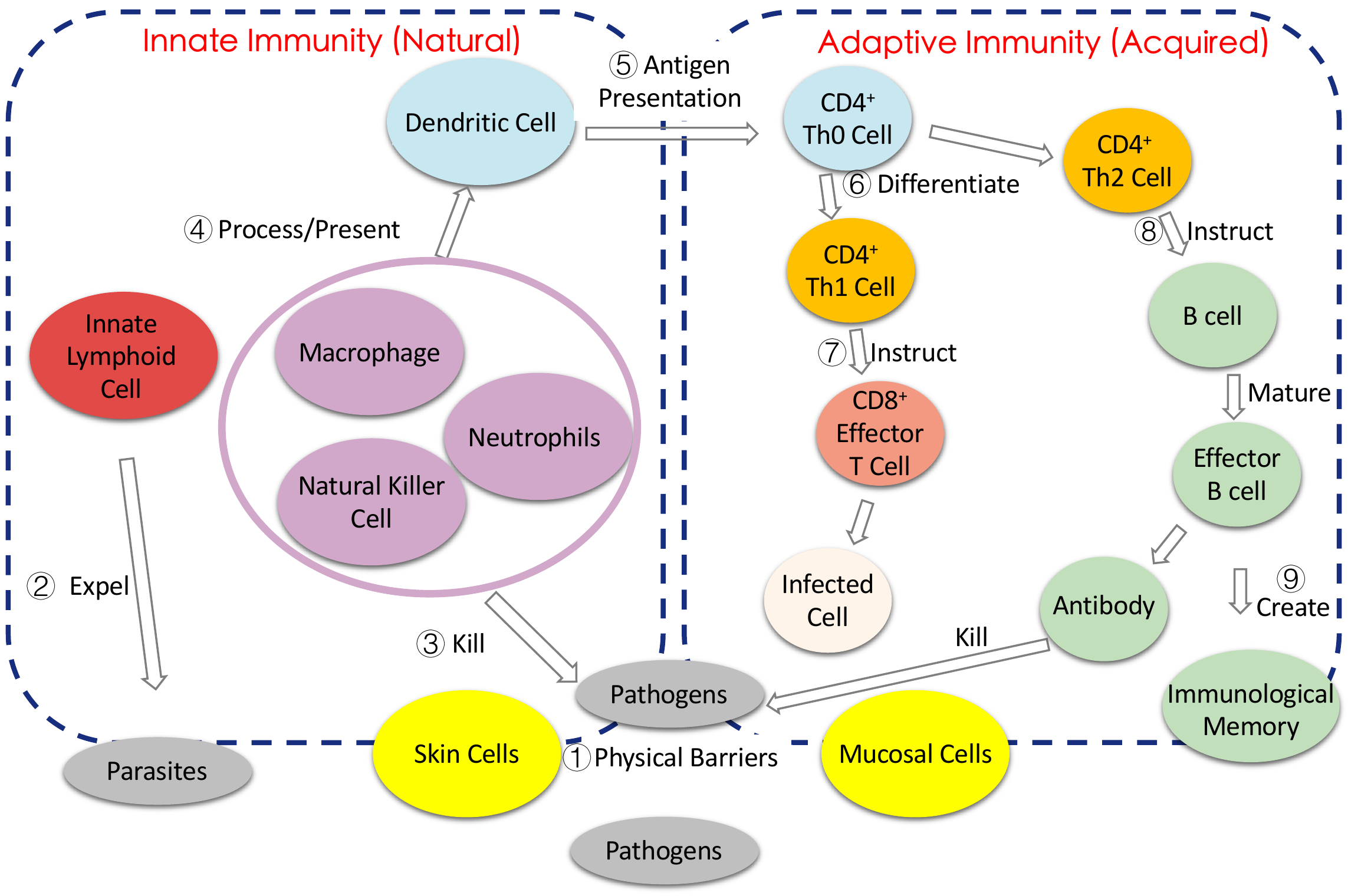}
\caption{The Basic Processes of the Immune System: the Innate and Adaptive Immune Systems}
\label{fig:biImmune-v2}
\end{figure*}

\subsection{The Basic Process of the Immune System}

 To achieve these primary functions, the immune system builds three lines of defense against invaders as shown in \figref{fig:biImmune-v2}. The basic processes of the immune system work as follows (the step number below corresponds to the number in Fig. 2).

\textbf{\uline{Step 1: Physical barriers prevent invaders from entering.}}

When the pathogens, such as viruses, bacteria, parasites, and fungi, try to invade the body, they meet the first defensive line, i.e., the physical barriers, including skins and mucous membrane \cite{how2019immune}. 
If some pathogens occasionally break the physical barriers, they will encounter the resistance from the innate and adaptive immune systems that is the second and third lines of defense. 

\textbf{\uline{Step 2: The innate immune system reacts immediately and expels parasites.}}

After infection, the body immediately activates the innate immune system to attack pathogens. 
Some large-size invaders, such as parasites, are excluded with the help of innate lymphoid cells (ILC). 

% \begin{figure*}[!t]
% \centering
% \includegraphics[width = 6in]{fig/bi-immune.pdf}
% \caption{The Basic Process of Immune System}
% \label{fig:biImmune}
% \end{figure*}

\textbf{\uline{Step 3: The innate immune cells attack the invaders.}}

Some white blood cells, e.g., natural killer (NK) cells, directly kill invaders by disrupting their cell walls. 
Some others, e.g., macrophages and neutrophils, are activated to ingest and destroy pathogens in a procedure called phagocytosis \cite{owen2013kuby}. 
Then we have an “inflammatory” response.

\textbf{\uline{Step 4: The innate immune cells extract features of the invaders.}}

Meanwhile, antigen presenting cells (APCs), e.g., macrophages and dendritic cells, extract features of pathogens through turning pathogenic proteins into peptides,  also called antigens. 
Later, they can present these antigens on their membrane surfaces to help adaptive immune cells, such as T cells, to recognize a specific type of pathogen \cite{owen2013kuby}.
This procedure is called \textbf{immune recognition}, and is part of the adaptive immune system. 

\textbf{\uline{Step 5: The adaptive immune system is activated to fight against invaders.}}

The antigen presented on the surface of APCs can activate the adaptive immune system, which can adapt to attack the specific invaders presented by this antigen.
It does this by activating lymphocytes, such as naive helper T cells. 

To fully activate naive T cells, APCs (typically dendritic cells) not only need to present antigens of the invaders but also need to release a second signal called costimulatory molecules. 
With these costimulatory molecules, the body alerts the adaptive immune system and recruits more lymphocytes to take part in the battle.

\textbf{\uline{Step 6: The adaptive immune system has two immune pathways to destroy infected cells and invaders.}}

After being activated, the naive helper T cells, i.e., CD4$^+$ Th0 cells, differentiate into two distinct functional subgroups, CD4$^+$ Th1 cells and CD4$^+$ Th2 cells, each of which produces a different set of cytokines. 
The CD4$^+$ Th1 cells start the pathway to destroy infected cells (called cell-mediated immunity), while the CD4$^+$ TH2 cells begin another pathway by which antibodies are produced to attack invaders (called humoral immunity).
In some particular cases, this differentiation can be done without the help of CD4$^+$.

\textbf{\uline{Step 7: The adaptive immune cells clear infected cells.}}

By excreting cytokines, CD4$^+$ Th1 helpers instruct the activated CD8$^+$ effector T cells to attack and clear the infected cells. 

\textbf{\uline{Step 8: The adaptive immune cells attack the invaders.}}

CD4$^+$ Th2 cells help B cells to mature into effector B cells that can produce antibodies that can bind to antigens in the pathogens in order to clear the pathogens in 7 to 10 days after infection \cite{owen2013kuby}. 
All these procedures described above are called the \textbf{immune response}.

\begin{figure*}[!t]
\centering
\includegraphics[width = 6in]{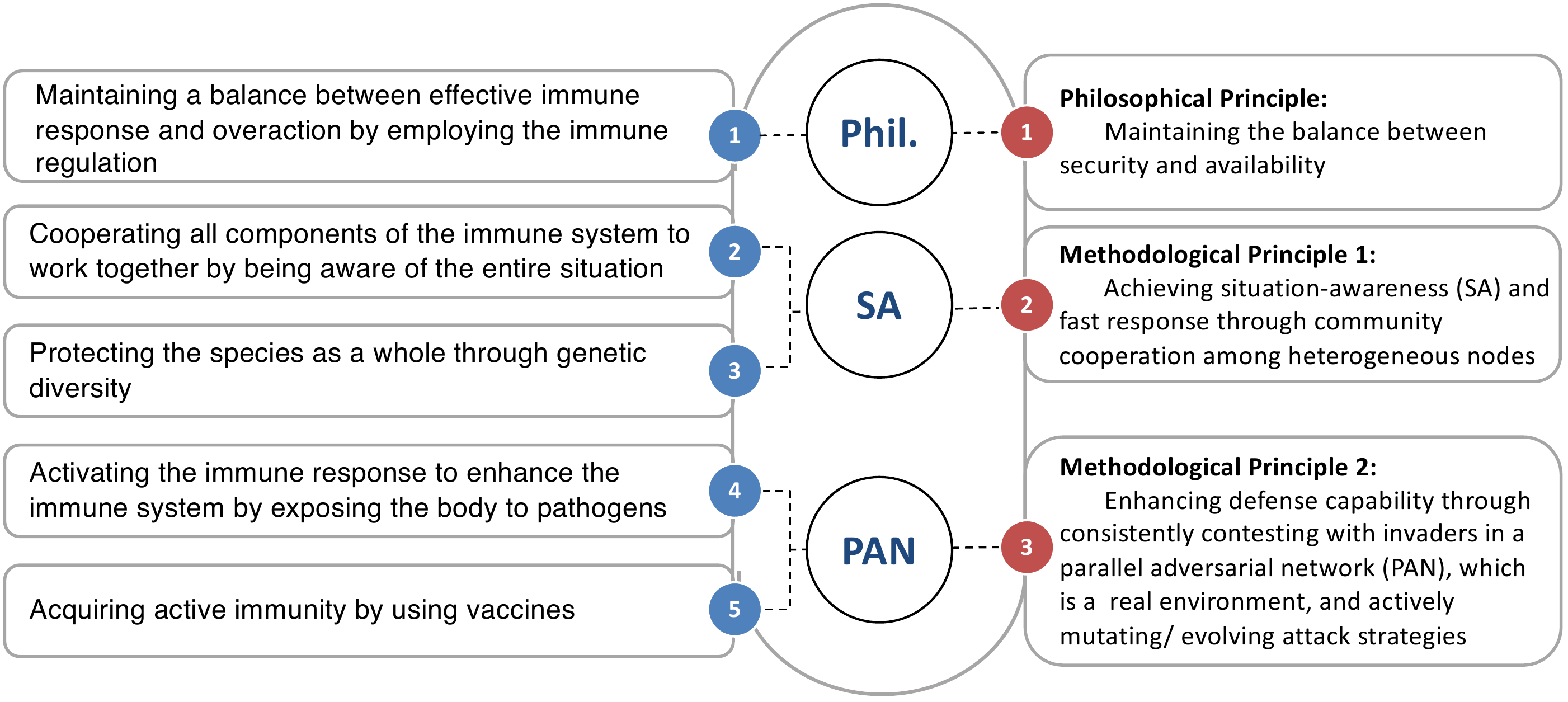}
\caption{The Inspirations and Principles Derived from The Immune System.}
\label{fig:principle}
\end{figure*}

\textbf{\uline{Step 9: The adaptive immune system produces the memory of the invaders for future use.}}

The helper T cells also help to produce memory lymphocytes. These lymphocytes can be activated more quickly than naive lymphocytes and can make a faster response when the same pathogens are encountered next time. 

Throughout this process, the immune system keeps an eye on the active level of invaders and the effects of the immune response. It carefully excretes positive regulators and negative regulators to \textbf{maintain the balance between effective immune response and overaction}, which called \textbf{immune regulation} (\figref{fig:Functions}). 

The whole process is repeated over and over again every day to defend the body. 
\textbf{By exposing to pathogens, the immune system is continuously activated, adjusted, and enforced to provide better protection}.

\subsection{Advanced Features of the Immune System}
In most cases, the individual immune system can provide adequate protection to us. However, it is not perfect. 
For example, the body needs 7 to 10 days to activate the adaptive immune system and what the immune system can do is limited when encountering virulent viruses, e.g., AIDS, Ebola and SARS. 
Thus,  the immune system evolves some intelligent methods, such as cooperation among different components, genetic diversity and vaccine, to deal with these issues.

\subsubsection{\textbf{Cooperation among Different Components of The Immune System}}
The immune system consists  various types of components, including immune cells and immune organs. 
Coordination between all these components to defend against pathogens requires the ability to know what happens, how serious it is, and how to recruit helpers.

To know what happens, the immune system places the innate immune cells throughout the body to observe pathogen invasions in the first place. 
They can also collect the features of the invaders and the location information about the invasion. 
Meanwhile, these innate immune cells and the non-immune cells at infection sites can excrete cytokines which indicate the threat level of the invasion. 
The immune system also designs various signaling pathways to deliver the threat information and uses the lymphatic system to transport immune cells to the battlefield effectively. 
With all of these mechanisms, the body can achieve situation-awareness and cooperation among different components of the immune system.

\subsubsection{\textbf{Genetic Diversity}}

As mentioned above, the immune system from one individual may not be good enough to eliminate unknown virulent viruses that are continually evolving. 
However, the reason that an outbred species can be resistant to these virulent viruses is due to genetic diversity. 
Different individuals have different biological features. 
Therefore, they can present and recognize different kinds of antigens. 
In most cases, some individuals cannot develop an immune response to a given pathogen and are vulnerable to the infection. 
However, the genetic diversities can provide a wide range of presentable antigens and effective antibodies against the antigens, which ensure some members of a species to survive. 

\subsubsection{\textbf{Vaccine}}
It is often not sufficient to enhance the immune system by simply waiting to encounter pathogens because it takes a long time to activate the immune system and sometimes the immune system may fail to activate.  
In an advanced society, by producing vaccines in the biology labs and injecting or swallowing them to uninfected individuals, their bodies can obtain an active acquired immunity to a pathogen, which can effectively restrain the spread of the disease. 
The vaccines are not harmful but have some properties of an antigen so that they can induce an immune response to generate antibodies against the antigen. 
This process is called the herd immunity. 
The biologists can now produce a specific type of bacteria by artificially generating mutations, and then make a vaccine for it in advance. 

\begin{figure*}[!t]
\centering
\includegraphics[width = 5.3in]{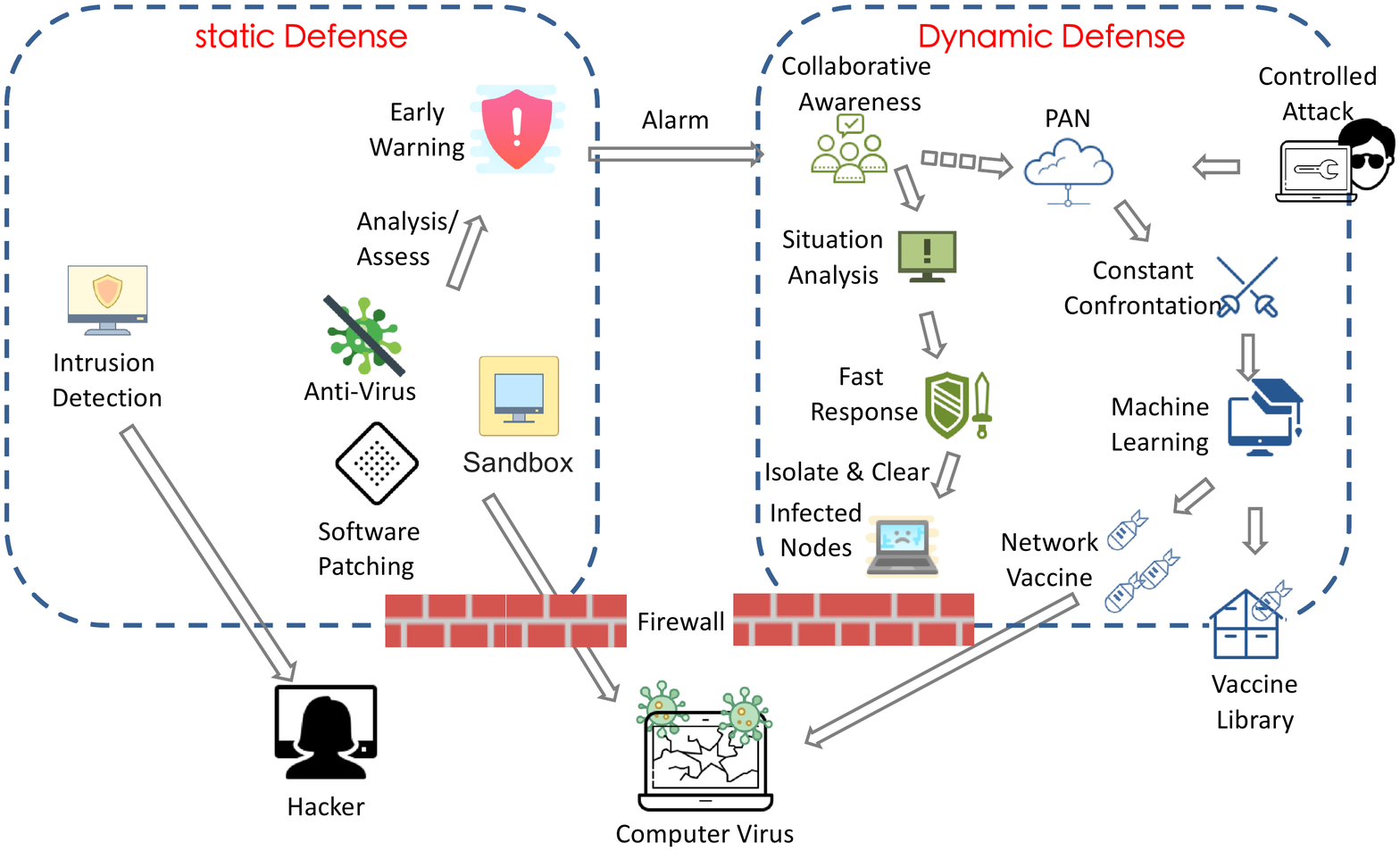}
\caption{The Reference Architecture for Network Security: Static and Dynamic defense}
\label{fig:arch}
\end{figure*}

\subsection{Inspirations Obtained from the Biological Immunology}
In summary, the body's immune system is always fighting against pathogens which can continue to mutate. 
To protect the body, the immune system endows several mechanisms which can inspire us on the design of the network security framework.  
 The most important inspirations we can learn from the biological immune system are listed below.
\begin{itemize}
\item Maintaining a balance between effective immune response and overaction by employing the immune regulation;
\item Cooperating all components of the immune system to work together by being aware of the entire situation;
\item Protecting the species as a whole through genetic diversity;
\item Activating the immune response to enhance the immune system by exposing the body to pathogens;
\item Acquiring active immunity by using vaccines.

\end{itemize}

\section*{A Network Security Architecture Inspired by Biological Immunology}
\label{sec.arch}

Inspired by the biological immune system, we propose a network security architecture. 
We first introduce the design principles and then present a reference architecture. 
%It is worth note that here we take network nodes analogous to individual organisms, and take the cyberspace analogous to biotic community. 

\setcounter{subsection}{0}
\subsection{Design Principles}

We  refine the five inspirations as three design principles, i.e., one philosophical principle and two methodological principles, for network security as shown in \figref{fig:principle}.

\ul{\textbf{- Philosophical Principle: maintaining the balance between security and availability.}}

Immune regulation, which balances effective immune response and overaction, is the most crucial feature of the biological immune system. Similarly, there should also be a ``trade-off" between network security and availability. 
Typically, to achieve absolute safety, some strict security measures, such as strict identity authentication, access control, data encryption, and digital signature, can be employed. 
However,  strict measures  make things difficult for ordinary users when they try to access the services. Thus, it is essential to maintainmaintain a balance between network security and availability. 
To achieve this goal, we present next two methodological principles.

\ul{\textbf{- Methodological Principle 1: Achieving situation awareness and fast response through community cooperation among heterogeneous nodes.}}

To achieve a balance between security and availability, it is critical to be aware of the entire situation (situation-awareness) about an impending invasion and share the situation among other nodes in real-time. 
With this situation information, these nodes can cooperatively evaluate the threat, issue a warning, and make a fast response to attackers. 
Meanwhile, the heterogeneity of nodes (e.g., using different operating systems and software) can help the entire system  survive when encountering unknown attacks. 

\ul{\textbf{- Methodological Principle 2: Enhancing defense capability through consistently contesting with invaders in a parallel adversarial network, which is a real environment, and actively mutating/evolving attack strategies.}}

% Another critical factor to keep balance is to learn attack and defend in advance, which means controllable real environment like a biology lab is needed. 
% With this environment, the network security architecture can launch adversarial attacks to observe, identify, evaluate, and assess network vulnerability while not affecting ordinary network traffic. 
% Further, to discover unknown attacks ahead of time, the network security architecture can allow the viruses and attack methods automatically mutate and evolve. Then, it can also employ techniques to develop methods automatically for dealing with these new threads. 

{Another critical factor to maintain a balance is to continue enhancing defense capability. By mimicking the human immune system, which is continuously exposed to pathogens, the network security architecture should have the ability to launch adversarial attacks to observe, identify, evaluate, and assess network vulnerability in a real environment{, which can be implemented as a parallel adversarial network (PAN), }without affecting ordinary network traffic. Further, to discover unknown attacks ahead of time, network security architecture should allow viruses and attack strategies to mutate and evolve automatically, like in a biology lab. Then, we can also employ techniques to develop defense tools automatically for dealing with these new threats.}

\subsection{Reference Framework}

 Following the design principles illustrated above, we propose a reference architecture, rather than a detailed design, for network security, as shown in \figref{fig:arch}. We intend to stress that this is a reference architecture for the entire cyberspace even though it looks like the individual immune system.   
This reference architecture demonstrates two kinds of functionalities a network security architecture should have, i.e., the functionalities of static defense and dynamic defense. 

\emph{1) \textbf{The Static Defense.}}

The static defense functionality provided by the network security architecture is the ability to react to invasions immediately and collect information about the invasion for dynamic defense.  
The network security architecture can achieve this functionality by deploying various defense methods, including firewalls, intrusion detection, software patching, sandboxes, and anti-virus softwares, everywhere. 
As these methods are located everywhere, they can observe invasions in the first place and generate a response quickly. 
They can also assess the threat level of the invasion and collect the features of the invaders to trigger an alarm to launch the dynamic defense.

\emph{2) \textbf{The Dynamic Defense.}}

The dynamic defense functionality is composed of the following two functions:

$\star$ \underline{Fast Response through Community Cooperation}

Following Methodological Principle 1, this function emphasizes the importance of cooperation and situation-awareness. 
Typically, there is only limited cooperation among different components of traditional network security architecture. 
However, being aware of the entire situation is the essential factor to fight against invaders.
Each component collects information, including network status, features of the attack, and the influence of the attack, and exchange this information through a signaling channel which is much faster than the data channel. 
This process can achieve collaborative awareness about the situation. 
With the awareness of the situation, the network security architecture can analyze the threat level of the invasion and make a fast response to isolate and clear the infected nodes.

$\star$ \ul{Acquiring adaptive response capabilities through adversarial attacks.}

As discussed in Methodological Principle 2, we need a controllable real environment to learn attack in advance to acquire the capabilities of adaptive responses without affecting ordinary network traffic.
Thus, we propose a parallel adversarial network (PAN) which accompanies the operation network. 
The PAN and the operation network share the same physical network resources by using network slicing techniques. 
In the PAN, we can proactively introduce controllable attack and viruses. 
We can also use machine learning methods to make attack and viruses mutate automatically. 
By doing this, we can observe, identify, evaluate, and assess network vulnerability.
Ultimately, we can produce network vaccines for defense, similar to vaccines in the immune system.

{The network security architecture that we proposed does not sacrifice availability, universality, and openness for security.  It achieves security by coordinating different components in the network to be cognizant of the entire situation for a fast response rather than deploying strict static security strategies. It also emphasizes to learn adversarial attacks in the parallel adversarial network, which is a real Internet as it shares the same physical network resources with the operational network by using network slicing.
Unlike the previous works of the artificial immune system that only mimic the methods of how the biological immune system distinguishes self and non-self cells to design new abnormal detection measures \cite{harmer2002artificial}, we also emphasize the importance of mutating and evolving attack strategies in a controllable real environment to produce network vaccines for proactive defense.}

\section*{Conclusion \& Future Work}
\label{sec.challenge}

In this article, we have briefly reviewed the biological immunology and acquired five inspirations from it.  
Base on these inspirations, we have proposed three design principles for a novel network security architecture.  
We have also presented a reference architecture design following the proposed principles.

To achieve a safer and more scalable architecture, there are still some research issues to be addressed in the future.

\subsubsection{Platform for Situation Awareness}
The goal of the platform for situation awareness is to provide a real-time sharing of global network situations to help nodes predict, analysis, and react to attacks in a cooperative and real-time manner. 
The critical issue is to build  such a platform that can deliver the situations and antigens much faster than  attackers. 

\subsubsection{Disclosing Situation Information}
Another critical issue for situation awareness is which information is essential and sufficient for situation awareness and how to encourage users to disclose this information.  This issue involves many considerations, such as personal privacy, techniques, and policy.  

\subsubsection{Parallel Adversarial Network (PAN)}
The parallel adversarial network is a real environment for performing adversarial attacks deliberately and obtaining network vaccine in advance. 
The most crucial issue is how to construct the PAN, make the PAN scale flexibly, and render the operation network impervious to the adversarial attacks performed in the PAN.  

\subsubsection{Conducting Adversarial Attacks}
Another important issue about the PAN is how to conduct adversarial attacks. We  need both the ability to replay an attack in the real network and the capability to mutate computer viruses and attacks automatically to conduct new attacks and generate new defense methods.

% \subsubsection{The criterion of balance}
% The biological immune system employs immune regulation to keep a balance between effective immune response and overaction. 
% It is crucial to investigate the criterion of balance and the methods to assess the situation.

\normalem

%\bibliographystyle{IEEEtranBST/IEEEtran}
%\normalem \bibliography{bib/bib}

\begin{thebibliography}{1}

\bibitem{forecast2019cisco}
Cisco,``Ciscovisualnetworkingindex:Forecastandtrends,2017–2022'' {\em White paper, Cisco Public Information}, 2019.

\bibitem{blumenthal2001rethinking}
M. S. Blumenthal and D. D. Clark, ``Rethinking the design of the internet: the end-to-end arguments vs. the brave new world,'' {\em ACM Transactions on Internet Technology (TOIT)}, vol. 1, no. 1, pp. 70–109, 2001.

\bibitem{james2017computer}
J. F. Kurose and K. W. Ross, ``Computer Networking: A top-down approach (7th Edition),'' {\em Pearson/Addison Wesley}, 2017.

\bibitem{US2019report}
The Council of Economic Advisers, ``The cost of malicious cyber activity to the u.s. economy,'' {\em White paper, the United States government}, 2018.

\bibitem{khan2018iot}
M. A. Khan and K. Salah, ``IoT security: Review, blockchain solutions, and open challenges,'' {\em Future Generation Computer Systems}, vol. 82, pp. 395–411, 2018.


\bibitem{spitzner2003honeynet}
L. Spitzner, ``The honeynet project: Trapping the hackers,'' {\em IEEE Security \& Privacy}, vol. 1, no. 2, pp. 15–23, 2003.

\bibitem{ferguson2014national}
B. Ferguson, A. Tall, and D. Olsen, ``National cyber range overview,'' in {\em Proc. 2014 IEEE Military Communications Conference}, 2014, pp. 123–128.

\bibitem{jajodia2011moving}
S. Jajodia, A. K. Ghosh, V. Swarup, C. Wang, and X. S. Wang, ``Moving target defense: creating asymmetric uncertainty for cyber threats,'' {\em Springer Science \& Business Media}, 2011, vol. 54.

\bibitem{mitchell2005trusted}
C. Mitchell, ``Trusted computing,'' {\em IET}, 2005.

\bibitem{hu2017mimic}
H. Hu, J. Wu, Z. Wang, and G. Cheng, ``Mimic defense: a designed-in cybersecurity defense framework,'' {\em IET Information Security}, vol. 12, no. 3, pp. 226–237, 2017.


\bibitem{abbas2014cellular}
A. K. Abbas, A. H. Lichtman, and S. Pillai, ``Cellular and molecular immunology E-book,'' {\em Elsevier Health Sciences}, 2014.

\bibitem{owen2013kuby}
 J. A. Owen, J. Punt, S. A. Stranford et al., ``Kuby immunology (7th Edition),'' {\em W.H. Freeman and Company}, 2013.

 \bibitem{immuneRegulation2007}
 A. Yoshimura, T. Naka, and M. Kubo, ``Socs proteins, cytokine sig- nalling and immune regulation,'' {\em Nature Reviews Immunology}, vol. 7, no. 6, p. 454, 2007.

 \bibitem{how2019immune}
 L. M. Sompayrac, ``How the immune system works (6th Edition),'' {\em Wiley-Blackwell}, 2019.
 
 \bibitem{harmer2002artificial}
 	P. K. Harmer, P. D. Williams, G. H. Gunsch, and G. B. Lamont, ``An artificial immune system architecture for computer security applications,'' {\em IEEE transactions on evolutionary computation}, vol. 6, no. 3, pp. 252–280, 2002.

\end{thebibliography}

\end{document}